\documentclass[preprint,authoryear,review,12pt]{elsarticle}

\usepackage{amssymb}

\usepackage{lineno}

\journal{Icarus}

\begin{document}

\newcommand{\cratio}{$^{12}$C/$^{13}$C}
\newcommand{\invcratio}{$^{13}$C/$^{12}$C}
\newcommand{\nratio}{$^{14}$N/$^{15}$N}
\newcommand{\oratio}{$^{16}$O/$^{18}$O}
\newcommand{\arratio}{$^{36}$Ar/$^{40}$Ar}

\newcommand{\methane}{CH$_4$}
\newcommand{\dmethane}{CH$_3$D}
\newcommand{\acetonitrile}{CH$_3$CN}
\newcommand{\ethane}{C$_2$H$_6$}
\newcommand{\acet}{C$_2$H$_2$}
\newcommand{\diacet}{C$_4$H$_2$}
\newcommand{\cyanoacet}{HC$_3$N}
\newcommand{\coo}{CO$_2$}
\newcommand{\hydrogen}{H$_2$}
\newcommand{\nitrogen}{N$_2$}
\newcommand{\ethynyl}{C$_2$H}
\newcommand{\methyl}{CH$_3$}
\newcommand{\methylene}{CH$_2$}
\newcommand{\ethylene}{C$_2$H$_4$}
\newcommand{\propane}{C$_3$H$_8$}
\newcommand{\phosphine}{PH$_3$}
\newcommand{\htwos}{H$_2$S}

\newcommand{\micron}{$\mu$m}
\newcommand{\cm}{cm$^{-1}$}
\newcommand{\dg}{$^{\circ}$}
\newcommand{\radunit}{W~cm$^{-2}\:$sr$^{-1}$/cm$^{-1}$}

\begin{frontmatter}



\title{Upper limits for \phosphine\ and \htwos\ in Titan's atmosphere from Cassini CIRS}


\author[gsfc]{Conor A. Nixon}
\address[gsfc]{Planetary Systems Laboratory, NASA Goddard Space Flight Center, Greenbelt, MD 20771, USA}
\author[bristol]{Nicholas A. Teanby}
\address[bristol]{School of Earth Sciences, University of Bristol, Wills Memorial Building, Queen's Road, BS8 1RJ, Bristol, UK}
\author[aopp]{Patrick G. J. Irwin}
\address[aopp]{Atmospheric, Oceanic and Planetary Physics, University of Oxford, Parks Road, Oxford, OX1 3PU, UK}
\author[cires]{Sarah M. H\"{o}rst}
\address[cires]{Cooperative Institute for Research in Environmental Sciences, University of Colorado, Boulder, CO 80309, USA}

\begin{abstract}
We have searched for the presence of simple P and S-bearing molecules in Titan's atmosphere, by looking for the characteristic signatures  of phosphine and hydrogen sulfide in infrared spectra obtained by Cassini CIRS. As a result we have placed the first upper limits on the stratospheric abundances, which are 1 ppb (\phosphine ) and 330 ppb  (\htwos ), at the 2-$\sigma$ significance level.

\end{abstract}


\end{frontmatter}

\linenumbers

\section{Introduction}
\label{sect:intro}

To date, no molecules bearing the light elements phosphorus ({P}) and  sulfur (S) have been detected in the atmosphere of Titan by either remote sensing or in-situ methods. However, from cosmological considerations both P and S must have been present in the icy planetesimals that formed Titan, and also delivered in trace quantities by later impacts. P in the form of \phosphine\ is found in Jupiter's atmosphere at approximately solar abundance, while on Saturn its abundance is around 3$\times$ solar \citep{owen03}. This enrichment is in line with core-accretion models, which predict that Saturn had a larger ice-to-gas fraction in its formation compared to Jupiter. Sulfur is enriched on Jupiter by 2.5$\times$ solar \citep{niemann98b} (and not yet detected on Saturn), perhaps due to easy formation of \htwos\ clathrate hydrates \citet{owen03}. Both P and S should therefore be present in Titan's bulk composition at fractions greater than those on Saturn, since the ice to gas ratio must have been much higher.

\citet{fortes07} considered a possible role for sulfur compounds in Titan cryovolcanism, suggesting that ammonium sulfate could form a magma in Titan's mantle. Plumes of this magma could dissolve methane present in crustal clathrates, allowing explosive release to the surface. One prediction of this model is that ammonium or other sulfates should be detectable spectroscopically on Titan's surface, as is the case on Europa and Ganymede. Such a model would presumably also release trace amounts of sulfur into Titan's atmosphere, especially in the wake of eruptions. \citet{pasek11} focused on the role of phosphorus in Titan, arguing that it could be delivered both endogenously and exogenously to the surface. Phosphine is efficiently trapped in clathrate hydrates, and \citet{pasek11} show that all clathrates that trap \htwos\ also trap at least as much \phosphine . Moreover, they show that \phosphine\ is more soluble in organic liquids than in water, and therefore any phosphine released from melted clathrates could be dissolved in Titan's hydrocarbon lakes, and would participate in the hydrocarbon meteorological cycle.

We have therefore searched the infrared spectrum of Titan for the signatures of the two most likely carriers of P and S: phosphine (\phosphine ) and hydrogen sulfide (\htwos ). Our data is from the Composite Infrared Spectrometer \citep[CIRS,][]{flasar04b}, carried onboard the Cassini spacecraft, which has been making close flybys of Titan since attaining Saturn orbit in mid-2004. Both gases have signatures in the 8--11 \micron\ (1250--900~\cm ) range, which is largely free of the hydrocarbon emissions that dominate much of Titan's infrared spectrum. We first model and subtract out the emissions of known species including methane (\methane ), acetylene (\acet ), ethylene (\ethylene ) and deuterated methane (\dmethane ). We then add \htwos\ and \phosphine\ incrementally to our spectral calculation using standard line lists from HITRAN \citep{rothman09}, and compare the predictions to the remaining Titan spectrum.

By comparing the model predictions to the instrument noise level at the 1, 2, and 3-$\sigma$ levels, we have derived the first numerical upper limits on the prevalence of \htwos\ and \phosphine\ in Titan's stratosphere. We follow a report of our results with a discussion of the implications for existing models, and finish with some conclusions about the prospects for future searches.

\section{Method}

\subsection{Instrument and Dataset}

The Cassini CIRS instrument is a dual design, comprising mid-infrared (1400--600~\cm ) and far-infrared (600--10~\cm ) spectrometers, both with a FWHM (full-width to half-maximum) spectral resolution variable from 0.5--15.5~\cm\  after Hamming apodization. The mid-IR Michelson spectrometer employs two $1\times10$ detector arrays: focal plane 3 (FP3, 600--1100~\cm ) and focal plane 4 (FP4, 1100--1400~\cm ). Both arrays have square pixels with fields of view 0.3 mrad across. FP4 has the highest sensitivity (lowest background noise) and least instrumental interferences, and was consequently used in this study.

The observations were described in an earlier paper \citep{nixon10b}. The spectra were acquired during the 55$^{th}$ Titan flyby (T55) of Cassini on May 22$^{nd}$ 2009, in a four-hour period when the spacecraft was at a range of 116,000--177,000~km. The observation (known as `MIRLMPAIR'-type) was targeted at Titan's limb, with the two arrays above and parallel to the horizon. FP4 was above FP3, with the pixels centered at an altitude of 247 km (0.27 mbar) and spanning approximately 44~km (just under one scale height) in the vertical direction at the mid-point of the observation. Using CIRS PAIR mode, all ten detectors simultaneously recorded data, paired into five receiver channels. A total of 941 pair-mode spectra at 0.5~\cm\ resolution were selected and then averaged to create a single, high signal-to-noise (S/N) ratio spectrum.

\subsection{Spectral Modeling}

The CIRS FP4 limb spectrum was then modeled to remove the signatures of known gas species. These included \methane\ ($\nu_4$ band at 1305~\cm ) and \dmethane\ ($\nu_6$ band at 1156~\cm ), plus some weaker contributions from the hydrocarbons \ethylene\ and \acet . The modeling closely follows that described in a recent paper \citep{nixon12b}, and is briefly summarized here.

An initial vertical atmospheric model was created with 100 layers equally spaced in log pressure from 1.45 bar to 0.05 $\mu$bar, based on the temperature profile and major gas abundances (\nitrogen, \methane , \hydrogen) determined by the Huygens probe \citep{fulchignoni05,niemann10}. Other gas species (\ethylene , \acet ) were included with constant vertical abundances in the stratosphere, at initial values taken from previous CIRS measurements at low latitudes \citep{coustenis10}. Similarly a uniformly mixed (constant particles/g atmosphere) stratospheric haze was included with optical properties from \citet{khare84}. The forward radiative transfer model was computed using the NEMESIS code \citep{irwin08} applied to the model atmosphere and using the HITRAN gas line atlas \citep{rothman09}, and then convolved with the FP4 detector spatial response shapes \citep{nixon09a} to generate a model spectrum.

\begin{figure}
\includegraphics[angle=0,scale=.65]{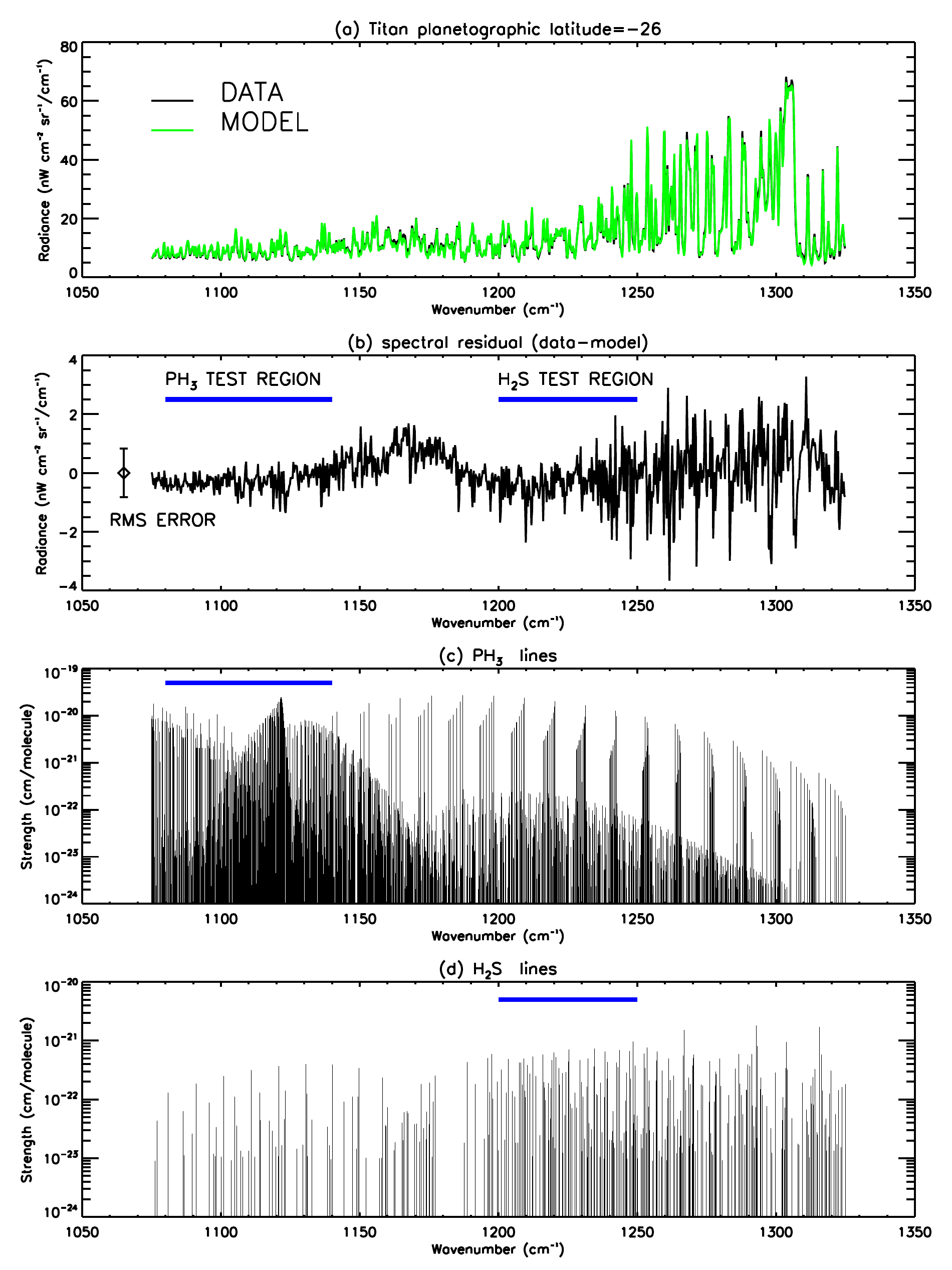}
\caption{(a): CIRS limb average spectrum of Titan from T55 flyby (black) and model fit (green). (b): data-model spectral residual. Note the $\nu_7$ band of \propane\ at 1158~\cm\ that is not included in our model (see text for details). ({c}) and (d) show the density and strengths of lines of \phosphine\ and \htwos\ respectively in the interval considered.}
\label{fig:fit}
\end{figure}

At this point the model was iterated to arrive at an optimum fit to the measured spectrum, by adjusting the model temperature profile (at each layer) and uniform vertical gas abundances of \acet\ and \ethylene , to minimize a cost function similar to a $\chi^2$ figure of merit. This approach has been successfully used to fit the T55 limb spectrum of Titan in previous work \citep{nixon09b,nixon10b}. Fig~\ref{fig:fit} (a) shows the best fit to the data. Note that a weak band of propane ($\nu_7$ at 1158~\cm ) was not included in the model, as a line list for this band has yet to be produced, and shows clearly in the data-model residual (Fig~\ref{fig:fit} (b)). The strong \methane\ $\nu_4$ band is fitted reasonably well, but also shows some residual mismatch, which may be due to imperfect knowledge of the underlying haze (continuum) opacity and/or wavelength calibration uncertainties. 

\subsection{Determination of Abundance Upper Limits}

To search for \phosphine\ and \htwos, we avoided spectral regions that showed non-random `noise' residual: especially the \propane\ $\nu_7$ region (1140--1200~\cm ) and the \methane\ region ($\tilde{\nu} > 1250$~\cm ). This permitted us to search for the $\nu_4$ band of \phosphine\ around 1120~\cm , and for a portion of the weak \htwos\ $\nu_2$ band at 1200--1250~\cm\ (see Fig~\ref{fig:fit} ({c}) and (d)). The summed line strengths in these regions were $1.40\times 10^{-18}$ cm molecule$^{-1}$ for 1121 lines of \phosphine\ from 1080--1140~\cm , and $2.2\times 10^{-20}$ cm molecule$^{-1}$ for 130 lines of \htwos\ at 1200--1250~\cm , using line data from HITRAN \citep{rothman09}. Note that the \phosphine\ band intensity was almost $100\times$ that of \htwos\ in the spectral ranges considered, so we expect $100\times$ greater sensitivity to \phosphine\ compared to \htwos .

\begin{figure}
\includegraphics[angle=0,scale=.65]{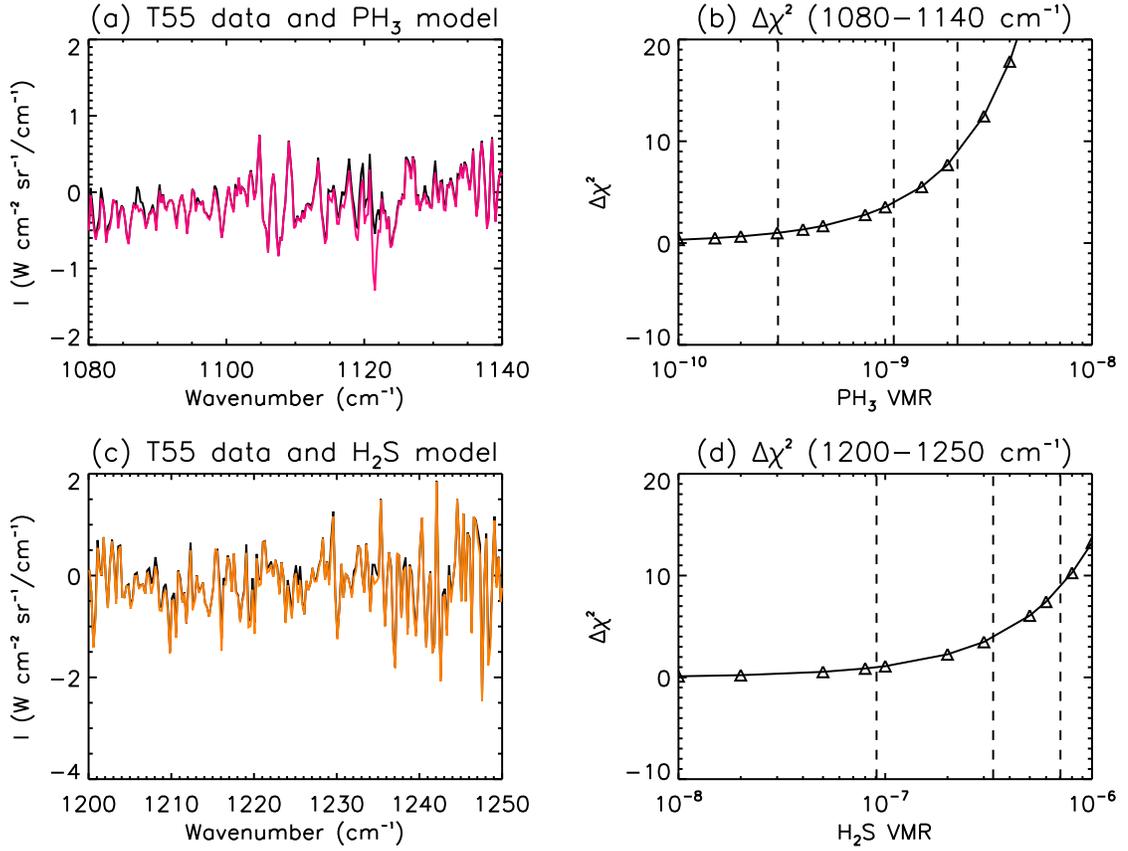}
\caption{Calculation of upper limits for the abundances of \htwos\ and \phosphine\ in Titan's stratosphere. (a) \& ({c}): residual of fit to the CIRS Titan spectrum in each region, after the modeling and removal known gas species (black line). Over-plotted is a calculation with exaggerated trial gas abundances of the undetected species to show their spectral signature (colored line). (b) \& (d): the curve of growth of $\Delta\chi^2$ over a wide range of trial abundances. The 1, 2 and 3-$\sigma$ upper limits are indicated by the vertical dashed lines at $\Delta\chi^2=$ 1, 4, 9 respectively.}
\label{fig:ulimits}
\end{figure}

These two gas species were added to the best-fit atmospheric model at fixed trial abundances of 1 ppb with other gases and temperature held constant at the previously retrieved values, and NEMESIS was allowed to retrieve a `best fit' abundance for the new species. In both cases, these retrievals resulted in no statistically significant improvement to the model $\chi^2$, hereafter designated $\chi_0^2$: the minimum $\chi^2$. We then proceeded to calculate upper limits to the abundances, following the approach of \citet{teanby09a} and \citet{nixon10b}. Starting with very low trial gas abundances in the model, these were increased incrementally, and at each value the forward spectral model was calculated, without optimized fitting or inversion, and the data-model $\chi^2$ computed. As the trial gas abundances increase, so too does the $\Delta\chi^2 = \chi^2 - \chi_0^2$. See Fig.~\ref{fig:ulimits}. Following \citet{press92} the 1, 2 and 3-$\sigma$ upper limits to the gas abundances occur at $\Delta\chi^2 = 1, 4, 9$.

\section{Results and Discussion}

Table 1 shows our results. The derived 1, 2, 3-$\sigma$ maximum abundances for \phosphine\ are 0.3, 1.1, 2.2 ppb respectively, while the corresponding values for \htwos\ are 91, 330, 700 ppb. The \htwos\ upper limits are some two orders of magnitude higher than those of \phosphine\ as expected, due to the much weaker band intensity in the spectral region considered, along with a higher 1-$\sigma$ NESR (Noise Equivalent Spectral Radiance) level in the corresponding spectral interval (0.6 versus 0.3 n\radunit ).

Sulfur compounds have been suggested to play a role in Titan cryovolcanism \citep{fortes07}, and therefore be present in trace amounts in Titan's atmosphere. Phosphorus is also easily dissolved in organic solvents, and as such could be a component of Titan's present-day methane-ethane lakes, and participate in the hydrocarbon cycle \citep{pasek11}. A simple calculation tracing the saturation vapor pressure (data from public NIST and CRC databases) up through Titan's troposphere using the Huygens temperature profile, indicates that abundances of \phosphine\ and \htwos\ that could reach the stratosphere are 6 and 0.035 ppb respectively. Therefore, our 2-$\sigma$ upper limit of 1 ppb for \phosphine\ provides some constraint on tropospheric \phosphine , while our \htwos\ limit of 330 ppb is four orders of magnitude greater than allowed in this scenario and is not constraining. 

At the present time, the lack of detection of P or S-bearing species anywhere in Titan's atmosphere has caused these elements to be omitted from present photochemical models, so we have no predictions for comparison. In future, it could be interesting for such models to allow for some injection of P and S molecules into the atmosphere through a putative eruption, and to investigate subsequent chemistry, for example, conversion of sulfates to sulfides. 

\section{Conclusions and Further Work}

In this paper we have searched for simple compounds of phosphorus and sulfur in Titan's infrared spectrum recorded by Cassini CIRS, placing the first upper limits on the abundance of \phosphine\ and \htwos . In the stratosphere at 247 km we find that no more than 1 ppb of \phosphine\ and 330 ppb of \htwos\ can be present, at the 2-$\sigma$ level of significance. Some potential exists for improving on these upper limits using CIRS, e.g. by future measurements at a lower limb altitude where the atmospheric density is greater. The peak sensitivity for weak trace gases is often near 10~mbar ($\sim$100~km), at least at low latitudes, so a repeat measurement in the lower stratosphere may yield more stringent limits.

Searching in the far-IR spectrum may prove helpful in the case of  \htwos , which has rotational lines at 50--150~\cm\  that are some $100\times $ stronger than the $\nu_2$ considered here. However in this region the NESR of CIRS FP1 is more than $10\times$ higher than that of the FP4 detectors used in this work, and suffers from systematic noise artifacts. Nevertheless, the sub-mm range has already proved productive for molecular line searches by instruments such as Herschel and ground-based sub-mm telescopes, resulting in the detections of \acetonitrile\  \citep{marten02} and HNC \citep{moreno11}, and will doubtless reveal further new species in due course.


\appendix

\clearpage

\begin{table}
{\bf Table 1: Calculated abundance upper limits for \phosphine\ and \htwos } \vspace*{2mm} \\
\begin{tabular}{lllllllll}
\hline
 & & Pressure & & Wavenumber & 1-$\sigma$ & \multicolumn{3}{c}{Upper Limits (ppb)} \\ \cline{7-9}
 Gas & Lat. & (mbar) & Band & Range (\cm ) & NESR$^{\ast}$ & 1-$\sigma$ & 2-$\sigma$ & 3-$\sigma$ \\
\hline
\phosphine\ & 25\dg S& 0.27 & $\nu_4$ & 1080--1140 & 0.29 & 0.30 & 1.1 & 2.2 \\
\htwos\ & 25\dg S & 0.27 & $\nu_2$ & 1200--1250 & 0.64 & 91 & 330 & 700 \\ 
\hline 
\end{tabular} \vspace*{2mm} \\
$^{\ast}$Noise Equivalent Spectral Radiance averaged over the spectral interval, in units of n\radunit .
\end{table}

\begin{thebibliography}{19}
\expandafter\ifx\csname natexlab\endcsname\relax\def\natexlab#1{#1}\fi
\expandafter\ifx\csname url\endcsname\relax
  \def\url#1{\texttt{#1}}\fi
\expandafter\ifx\csname urlprefix\endcsname\relax\def\urlprefix{URL }\fi
\providecommand{\eprint}[2][]{\url{#2}}
\providecommand{\bibinfo}[2]{#2}
\ifx\xfnm\relax \def\xfnm[#1]{\unskip,\space#1}\fi
\bibitem[{{Coustenis} et~al.(2010){Coustenis}, {Jennings}, {Nixon},
  {Achterberg}, {Lavvas}, {Vinatier}, {Teanby}, {Bjoraker}, {Carlson}, {Piani},
  {Bampasidis}, {Flasar} and {Romani}}]{coustenis10}
\bibinfo{author}{{Coustenis}, A.}, \bibinfo{author}{{Jennings}, D.E.},
  \bibinfo{author}{{Nixon}, C.A.}, \bibinfo{author}{{Achterberg}, R.K.},
  \bibinfo{author}{{Lavvas}, P.}, \bibinfo{author}{{Vinatier}, S.},
  \bibinfo{author}{{Teanby}, N.A.}, \bibinfo{author}{{Bjoraker}, G.L.},
  \bibinfo{author}{{Carlson}, R.C.}, \bibinfo{author}{{Piani}, L.},
  \bibinfo{author}{{Bampasidis}, G.}, \bibinfo{author}{{Flasar}, F.M.},
  \bibinfo{author}{{Romani}, P.N.}, \bibinfo{year}{2010}.
\newblock \bibinfo{title}{{Titan trace gaseous composition from CIRS at the end
  of the Cassini-Huygens prime mission}}.
\newblock \bibinfo{journal}{{Icarus}} \bibinfo{volume}{207},
  \bibinfo{pages}{461--476}.
\bibitem[{{Flasar} et~al.(2004){Flasar}, {Kunde}, {Abbas}, {Achterberg}, {Ade},
  {Barucci}, {B{\'e}zard}, {Bjoraker}, {Brasunas}, {Calcutt}, {Carlson},
  {C{\'e}sarsky}, {Conrath}, {Coradini}, {Courtin}, {Coustenis}, {Edberg},
  {Edgington}, {Ferrari}, {Fouchet}, {Gautier}, {Gierasch}, {Grossman},
  {Irwin}, {Jennings}, {Lellouch}, {Mamoutkine}, {Marten}, {Meyer}, {Nixon},
  {Orton}, {Owen}, {Pearl}, {Prang{\'e}}, {Raulin}, {Read}, {Romani},
  {Samuelson}, {Segura}, {Showalter}, {Simon-Miller}, {Smith}, {Spencer},
  {Spilker} and {Taylor}}]{flasar04b}
\bibinfo{author}{{Flasar}, F.M.}, \bibinfo{author}{{Kunde}, V.G.},
  \bibinfo{author}{{Abbas}, M.M.}, \bibinfo{author}{{Achterberg}, R.K.},
  \bibinfo{author}{{Ade}, P.}, \bibinfo{author}{{Barucci}, A.},
  \bibinfo{author}{{B{\'e}zard}, B.}, \bibinfo{author}{{Bjoraker}, G.L.},
  \bibinfo{author}{{Brasunas}, J.C.}, \bibinfo{author}{{Calcutt}, S.},
  \bibinfo{author}{{Carlson}, R.}, \bibinfo{author}{{C{\'e}sarsky}, C.J.},
  \bibinfo{author}{{Conrath}, B.J.}, \bibinfo{author}{{Coradini}, A.},
  \bibinfo{author}{{Courtin}, R.}, \bibinfo{author}{{Coustenis}, A.},
  \bibinfo{author}{{Edberg}, S.}, \bibinfo{author}{{Edgington}, S.},
  \bibinfo{author}{{Ferrari}, C.}, \bibinfo{author}{{Fouchet}, T.},
  \bibinfo{author}{{Gautier}, D.}, \bibinfo{author}{{Gierasch}, P.J.},
  \bibinfo{author}{{Grossman}, K.}, \bibinfo{author}{{Irwin}, P.},
  \bibinfo{author}{{Jennings}, D.E.}, \bibinfo{author}{{Lellouch}, E.},
  \bibinfo{author}{{Mamoutkine}, A.A.}, \bibinfo{author}{{Marten}, A.},
  \bibinfo{author}{{Meyer}, J.P.}, \bibinfo{author}{{Nixon}, C.A.},
  \bibinfo{author}{{Orton}, G.S.}, \bibinfo{author}{{Owen}, T.C.},
  \bibinfo{author}{{Pearl}, J.C.}, \bibinfo{author}{{Prang{\'e}}, R.},
  \bibinfo{author}{{Raulin}, F.}, \bibinfo{author}{{Read}, P.L.},
  \bibinfo{author}{{Romani}, P.N.}, \bibinfo{author}{{Samuelson}, R.E.},
  \bibinfo{author}{{Segura}, M.E.}, \bibinfo{author}{{Showalter}, M.R.},
  \bibinfo{author}{{Simon-Miller}, A.A.}, \bibinfo{author}{{Smith}, M.D.},
  \bibinfo{author}{{Spencer}, J.R.}, \bibinfo{author}{{Spilker}, L.J.},
  \bibinfo{author}{{Taylor}, F.W.}, \bibinfo{year}{2004}.
\newblock \bibinfo{title}{{Exploring The Saturn System In The Thermal Infrared:
  The Composite Infrared Spectrometer}}.
\newblock \bibinfo{journal}{{Space Sci. Rev.}} \bibinfo{volume}{115},
  \bibinfo{pages}{169--297}.
\bibitem[{{Fortes} et~al.(2007){Fortes}, {Grindrod}, {Trickett} and {Vo{\v
  c}adlo}}]{fortes07}
\bibinfo{author}{{Fortes}, A.D.}, \bibinfo{author}{{Grindrod}, P.M.},
  \bibinfo{author}{{Trickett}, S.K.}, \bibinfo{author}{{Vo{\v c}adlo}, L.},
  \bibinfo{year}{2007}.
\newblock \bibinfo{title}{{Ammonium sulfate on Titan: Possible origin and role
  in cryovolcanism}}.
\newblock \bibinfo{journal}{{Icarus}} \bibinfo{volume}{188},
  \bibinfo{pages}{139--153}.
\bibitem[{{Fulchignoni} et~al.(2005){Fulchignoni}, {Ferri}, {Angrilli}, {Ball},
  {Bar-Nun}, {Barucci}, {Bettanini}, {Bianchini}, {Borucki}, {Colombatti},
  {Coradini}, {Coustenis}, {Debei}, {Falkner}, {Fanti}, {Flamini}, {Gaborit},
  {Grard}, {Hamelin}, {Harri}, {Hathi}, {Jernej}, {Leese}, {Lehto}, {Lion
  Stoppato}, {L{\'o}pez-Moreno}, {M{\"a}kinen}, {McDonnell}, {McKay},
  {Molina-Cuberos}, {Neubauer}, {Pirronello}, {Rodrigo}, {Saggin},
  {Schwingenschuh}, {Seiff}, {Sim{\~o}es}, {Svedhem}, {Tokano}, {Towner},
  {Trautner}, {Withers} and {Zarnecki}}]{fulchignoni05}
\bibinfo{author}{{Fulchignoni}, M.}, \bibinfo{author}{{Ferri}, F.},
  \bibinfo{author}{{Angrilli}, F.}, \bibinfo{author}{{Ball}, A.J.},
  \bibinfo{author}{{Bar-Nun}, A.}, \bibinfo{author}{{Barucci}, M.A.},
  \bibinfo{author}{{Bettanini}, C.}, \bibinfo{author}{{Bianchini}, G.},
  \bibinfo{author}{{Borucki}, W.}, \bibinfo{author}{{Colombatti}, G.},
  \bibinfo{author}{{Coradini}, M.}, \bibinfo{author}{{Coustenis}, A.},
  \bibinfo{author}{{Debei}, S.}, \bibinfo{author}{{Falkner}, P.},
  \bibinfo{author}{{Fanti}, G.}, \bibinfo{author}{{Flamini}, E.},
  \bibinfo{author}{{Gaborit}, V.}, \bibinfo{author}{{Grard}, R.},
  \bibinfo{author}{{Hamelin}, M.}, \bibinfo{author}{{Harri}, A.M.},
  \bibinfo{author}{{Hathi}, B.}, \bibinfo{author}{{Jernej}, I.},
  \bibinfo{author}{{Leese}, M.R.}, \bibinfo{author}{{Lehto}, A.},
  \bibinfo{author}{{Lion Stoppato}, P.F.}, \bibinfo{author}{{L{\'o}pez-Moreno},
  J.J.}, \bibinfo{author}{{M{\"a}kinen}, T.}, \bibinfo{author}{{McDonnell},
  J.A.M.}, \bibinfo{author}{{McKay}, C.P.}, \bibinfo{author}{{Molina-Cuberos},
  G.}, \bibinfo{author}{{Neubauer}, F.M.}, \bibinfo{author}{{Pirronello}, V.},
  \bibinfo{author}{{Rodrigo}, R.}, \bibinfo{author}{{Saggin}, B.},
  \bibinfo{author}{{Schwingenschuh}, K.}, \bibinfo{author}{{Seiff}, A.},
  \bibinfo{author}{{Sim{\~o}es}, F.}, \bibinfo{author}{{Svedhem}, H.},
  \bibinfo{author}{{Tokano}, T.}, \bibinfo{author}{{Towner}, M.C.},
  \bibinfo{author}{{Trautner}, R.}, \bibinfo{author}{{Withers}, P.},
  \bibinfo{author}{{Zarnecki}, J.C.}, \bibinfo{year}{2005}.
\newblock \bibinfo{title}{{In situ measurements of the physical characteristics
  of Titan's environment}}.
\newblock \bibinfo{journal}{{Nature}} \bibinfo{volume}{438},
  \bibinfo{pages}{785--791}.
\bibitem[{{Irwin} et~al.(2008){Irwin}, {Teanby}, {de Kok}, {Fletcher},
  {Howett}, {Tsang}, {Wilson}, {Calcutt}, {Nixon} and {Parrish}}]{irwin08}
\bibinfo{author}{{Irwin}, P.G.J.}, \bibinfo{author}{{Teanby}, N.A.},
  \bibinfo{author}{{de Kok}, R.}, \bibinfo{author}{{Fletcher}, L.N.},
  \bibinfo{author}{{Howett}, C.J.A.}, \bibinfo{author}{{Tsang}, C.C.C.},
  \bibinfo{author}{{Wilson}, C.F.}, \bibinfo{author}{{Calcutt}, S.B.},
  \bibinfo{author}{{Nixon}, C.A.}, \bibinfo{author}{{Parrish}, P.D.},
  \bibinfo{year}{2008}.
\newblock \bibinfo{title}{{The NEMESIS planetary atmosphere radiative transfer
  and retrieval tool}}.
\newblock \bibinfo{journal}{{J. Quant. Spectr. Rad. Trans.}}
  \bibinfo{volume}{109}, \bibinfo{pages}{1136--1150}.
\bibitem[{{Khare} et~al.(1984){Khare}, {Sagan}, {Arakawa}, {Suits}, {Callcott}
  and {Williams}}]{khare84}
\bibinfo{author}{{Khare}, B.N.}, \bibinfo{author}{{Sagan}, C.},
  \bibinfo{author}{{Arakawa}, E.T.}, \bibinfo{author}{{Suits}, F.},
  \bibinfo{author}{{Callcott}, T.A.}, \bibinfo{author}{{Williams}, M.W.},
  \bibinfo{year}{1984}.
\newblock \bibinfo{title}{{Optical constants of organic tholins produced in a
  simulated Titanian atmosphere - From soft X-ray to microwave frequencies}}.
\newblock \bibinfo{journal}{{Icarus}} \bibinfo{volume}{60},
  \bibinfo{pages}{127--137}.
\bibitem[{{Marten} et~al.(2002){Marten}, {Hidayat}, {Biraud} and
  {Moreno}}]{marten02}
\bibinfo{author}{{Marten}, A.}, \bibinfo{author}{{Hidayat}, T.},
  \bibinfo{author}{{Biraud}, Y.}, \bibinfo{author}{{Moreno}, R.},
  \bibinfo{year}{2002}.
\newblock \bibinfo{title}{{New Millimeter Heterodyne Observations of Titan:
  Vertical Distributions of Nitriles HCN, HC$_3$N, CH$_{3}$CN, and the Isotopic
  Ratio $^{15}$N/$^{14}$N in Its Atmosphere}}.
\newblock \bibinfo{journal}{{Icarus}} \bibinfo{volume}{158},
  \bibinfo{pages}{532--544}.
\bibitem[{{Moreno} et~al.(2011){Moreno}, {Lellouch}, {Lara}, {Courtin},
  {Bockel{\'e}e-Morvan}, {Hartogh}, {Rengel}, {Biver}, {Banaszkiewicz} and
  {Gonz{\'a}lez}}]{moreno11}
\bibinfo{author}{{Moreno}, R.}, \bibinfo{author}{{Lellouch}, E.},
  \bibinfo{author}{{Lara}, L.M.}, \bibinfo{author}{{Courtin}, R.},
  \bibinfo{author}{{Bockel{\'e}e-Morvan}, D.}, \bibinfo{author}{{Hartogh}, P.},
  \bibinfo{author}{{Rengel}, M.}, \bibinfo{author}{{Biver}, N.},
  \bibinfo{author}{{Banaszkiewicz}, M.}, \bibinfo{author}{{Gonz{\'a}lez}, A.},
  \bibinfo{year}{2011}.
\newblock \bibinfo{title}{{First detection of hydrogen isocyanide (HNC) in
  Titan's atmosphere}}.
\newblock \bibinfo{journal}{{Astron. and Astrophys.}} \bibinfo{volume}{536},
  \bibinfo{pages}{L12}.
\bibitem[{{Niemann} et~al.(1998){Niemann}, {Atreya}, {Carignan}, {Donahue},
  {Haberman}, {Harpold}, {Hartle}, {Hunten}, {Kasprzak}, {Mahaffy}, {Owen} and
  {Way}}]{niemann98b}
\bibinfo{author}{{Niemann}, H.B.}, \bibinfo{author}{{Atreya}, S.K.},
  \bibinfo{author}{{Carignan}, G.R.}, \bibinfo{author}{{Donahue}, T.M.},
  \bibinfo{author}{{Haberman}, J.A.}, \bibinfo{author}{{Harpold}, D.N.},
  \bibinfo{author}{{Hartle}, R.E.}, \bibinfo{author}{{Hunten}, D.M.},
  \bibinfo{author}{{Kasprzak}, W.T.}, \bibinfo{author}{{Mahaffy}, P.R.},
  \bibinfo{author}{{Owen}, T.C.}, \bibinfo{author}{{Way}, S.H.},
  \bibinfo{year}{1998}.
\newblock \bibinfo{title}{{The composition of the Jovian atmosphere as
  determined by the Galileo probe mass spectrometer}}.
\newblock \bibinfo{journal}{{J. Geophys. Res.}} \bibinfo{volume}{103},
  \bibinfo{pages}{22831--22846}.
\bibitem[{{Niemann} et~al.(2010){Niemann}, {Atreya}, {Demick}, {Gautier},
  {Haberman}, {Harpold}, {Kasprzak}, {Lunine}, {Owen} and {Raulin}}]{niemann10}
\bibinfo{author}{{Niemann}, H.B.}, \bibinfo{author}{{Atreya}, S.K.},
  \bibinfo{author}{{Demick}, J.E.}, \bibinfo{author}{{Gautier}, D.},
  \bibinfo{author}{{Haberman}, J.A.}, \bibinfo{author}{{Harpold}, D.N.},
  \bibinfo{author}{{Kasprzak}, W.T.}, \bibinfo{author}{{Lunine}, J.I.},
  \bibinfo{author}{{Owen}, T.C.}, \bibinfo{author}{{Raulin}, F.},
  \bibinfo{year}{2010}.
\newblock \bibinfo{title}{{Composition of Titan's lower atmosphere and simple
  surface volatiles as measured by the Cassini-Huygens probe gas chromatograph
  mass spectrometer experiment}}.
\newblock \bibinfo{journal}{Journal of Geophysical Research (Planets)}
  \bibinfo{volume}{115}, \bibinfo{pages}{12006}.
\bibitem[{{Nixon} et~al.(2010){Nixon}, {Achterberg}, {Teanby}, {Irwin},
  {Flaud}, {Kleiner}, {Dehayem-Kamadjeu}, {Brown}, {Sams}, {B{\'e}zard},
  {Coustenis}, {Ansty}, {Mamoutkine}, {Vinatier}, {Bjoraker}, {Jennings},
  {Romani} and {Flasar}}]{nixon10b}
\bibinfo{author}{{Nixon}, C.A.}, \bibinfo{author}{{Achterberg}, R.K.},
  \bibinfo{author}{{Teanby}, N.A.}, \bibinfo{author}{{Irwin}, P.G.J.},
  \bibinfo{author}{{Flaud}, J.M.}, \bibinfo{author}{{Kleiner}, I.},
  \bibinfo{author}{{Dehayem-Kamadjeu}, A.}, \bibinfo{author}{{Brown}, L.R.},
  \bibinfo{author}{{Sams}, R.L.}, \bibinfo{author}{{B{\'e}zard}, B.},
  \bibinfo{author}{{Coustenis}, A.}, \bibinfo{author}{{Ansty}, T.M.},
  \bibinfo{author}{{Mamoutkine}, A.}, \bibinfo{author}{{Vinatier}, S.},
  \bibinfo{author}{{Bjoraker}, G.L.}, \bibinfo{author}{{Jennings}, D.E.},
  \bibinfo{author}{{Romani}, P.N.}, \bibinfo{author}{{Flasar}, F.M.},
  \bibinfo{year}{2010}.
\newblock \bibinfo{title}{{Upper limits for undetected trace species in the
  stratosphere of Titan}}.
\newblock \bibinfo{journal}{Faraday Discussions} \bibinfo{volume}{147},
  \bibinfo{pages}{65}.
\newblock \eprint{1103.0297}.
\bibitem[{{Nixon} et~al.(2009a){Nixon}, {Jennings}, {Flaud}, {B{\'e}zard},
  {Teanby}, {Irwin}, {Ansty}, {Coustenis}, {Vinatier} and {Flasar}}]{nixon09b}
\bibinfo{author}{{Nixon}, C.A.}, \bibinfo{author}{{Jennings}, D.E.},
  \bibinfo{author}{{Flaud}, J.M.}, \bibinfo{author}{{B{\'e}zard}, B.},
  \bibinfo{author}{{Teanby}, N.A.}, \bibinfo{author}{{Irwin}, P.G.J.},
  \bibinfo{author}{{Ansty}, T.M.}, \bibinfo{author}{{Coustenis}, A.},
  \bibinfo{author}{{Vinatier}, S.}, \bibinfo{author}{{Flasar}, F.M.},
  \bibinfo{year}{2009}a.
\newblock \bibinfo{title}{{Titan's prolific propane: The Cassini CIRS
  perspective}}.
\newblock \bibinfo{journal}{{Planetary and Space Science}}
  \bibinfo{volume}{57}, \bibinfo{pages}{1573--1585}.
\newblock \eprint{0909.1794}.
\bibitem[{{Nixon} et~al.(2009b){Nixon}, {Teanby}, {Calcutt}, {Aslam},
  {Jennings}, {Kunde}, {Flasar}, {Irwin}, {Taylor}, {Glenar} and
  {Smith}}]{nixon09a}
\bibinfo{author}{{Nixon}, C.A.}, \bibinfo{author}{{Teanby}, N.A.},
  \bibinfo{author}{{Calcutt}, S.B.}, \bibinfo{author}{{Aslam}, S.},
  \bibinfo{author}{{Jennings}, D.E.}, \bibinfo{author}{{Kunde}, V.G.},
  \bibinfo{author}{{Flasar}, F.M.}, \bibinfo{author}{{Irwin}, P.G.},
  \bibinfo{author}{{Taylor}, F.W.}, \bibinfo{author}{{Glenar}, D.A.},
  \bibinfo{author}{{Smith}, M.D.}, \bibinfo{year}{2009}b.
\newblock \bibinfo{title}{{Infrared limb sounding of Titan with the Cassini
  Composite InfraRed Spectrometer: effects of the mid-IR detector spatial
  responses}}.
\newblock \bibinfo{journal}{{Applied Optics}} \bibinfo{volume}{48},
  \bibinfo{pages}{1912}.
\bibitem[{{Nixon} et~al.(2012){Nixon}, {Temelso}, {Vinatier}, {Teanby},
  {B\'{e}zard}, {Achterberg}, {Mandt}, {Sherrill}, {Irwin}, {Jennings},
  {Romani}, {Coustenis} and {Flasar}}]{nixon12b}
\bibinfo{author}{{Nixon}, C.A.}, \bibinfo{author}{{Temelso}, B.},
  \bibinfo{author}{{Vinatier}, S.}, \bibinfo{author}{{Teanby}, N.A.},
  \bibinfo{author}{{B\'{e}zard}, B.}, \bibinfo{author}{{Achterberg}, R.K.},
  \bibinfo{author}{{Mandt}, K.E.}, \bibinfo{author}{{Sherrill}, C.D.},
  \bibinfo{author}{{Irwin}, P.G.J.}, \bibinfo{author}{{Jennings}, D.E.},
  \bibinfo{author}{{Romani}, P.N.}, \bibinfo{author}{{Coustenis}, A.},
  \bibinfo{author}{{Flasar}, F.M.}, \bibinfo{year}{2012}.
\newblock \bibinfo{title}{{Isotopic ratios in Titan's methane: measurements and
  modeling}}.
\newblock \bibinfo{journal}{{Astrophys. J.}} \bibinfo{volume}{749},
  \bibinfo{pages}{159}.
\bibitem[{{Owen} and {Encrenaz}(2003)}]{owen03}
\bibinfo{author}{{Owen}, T.}, \bibinfo{author}{{Encrenaz}, T.},
  \bibinfo{year}{2003}.
\newblock \bibinfo{title}{{Element Abundances and Isotope Ratios in the Giant
  Planets and Titan}}.
\newblock \bibinfo{journal}{{Space Sci. Rev.}} \bibinfo{volume}{106},
  \bibinfo{pages}{121--138}.
\bibitem[{{Pasek} et~al.(2011){Pasek}, {Mousis} and {Lunine}}]{pasek11}
\bibinfo{author}{{Pasek}, M.A.}, \bibinfo{author}{{Mousis}, O.},
  \bibinfo{author}{{Lunine}, J.I.}, \bibinfo{year}{2011}.
\newblock \bibinfo{title}{{Phosphorus chemistry on Titan}}.
\newblock \bibinfo{journal}{{Icarus}} \bibinfo{volume}{212},
  \bibinfo{pages}{751--761}.
\bibitem[{{Press} et~al.(1992){Press}, {Teukolsky}, {Vetterling} and
  {Flannery}}]{press92}
\bibinfo{author}{{Press}, W.H.}, \bibinfo{author}{{Teukolsky}, S.A.},
  \bibinfo{author}{{Vetterling}, W.T.}, \bibinfo{author}{{Flannery}, B.P.},
  \bibinfo{year}{1992}.
\newblock \bibinfo{title}{{Numerical recipes in FORTRAN. The art of scientific
  computing}}.
\bibitem[{{Rothman} et~al.(2009){Rothman}, {Gordon}, {Barbe}, {Benner},
  {Bernath}, {Birk}, {Boudon}, {Brown}, {Campargue}, {Champion}, {Chance},
  {Coudert}, {Dana}, {Devi}, {Fally}, {Flaud}, {Gamache}, {Goldman},
  {Jacquemart}, {Kleiner}, {Lacome}, {Lafferty}, {Mandin}, {Massie},
  {Mikhailenko}, {Miller}, {Moazzen-Ahmadi}, {Naumenko}, {Nikitin}, {Orphal},
  {Perevalov}, {Perrin}, {Predoi-Cross}, {Rinsland}, {Rotger}, {{\v S}ime{\v
  c}kov{\'a}}, {Smith}, {Sung}, {Tashkun}, {Tennyson}, {Toth}, {Vandaele} and
  {Vander Auwera}}]{rothman09}
\bibinfo{author}{{Rothman}, L.S.}, \bibinfo{author}{{Gordon}, I.E.},
  \bibinfo{author}{{Barbe}, A.}, \bibinfo{author}{{Benner}, D.C.},
  \bibinfo{author}{{Bernath}, P.F.}, \bibinfo{author}{{Birk}, M.},
  \bibinfo{author}{{Boudon}, V.}, \bibinfo{author}{{Brown}, L.R.},
  \bibinfo{author}{{Campargue}, A.}, \bibinfo{author}{{Champion}, J.P.},
  \bibinfo{author}{{Chance}, K.}, \bibinfo{author}{{Coudert}, L.H.},
  \bibinfo{author}{{Dana}, V.}, \bibinfo{author}{{Devi}, V.M.},
  \bibinfo{author}{{Fally}, S.}, \bibinfo{author}{{Flaud}, J.M.},
  \bibinfo{author}{{Gamache}, R.R.}, \bibinfo{author}{{Goldman}, A.},
  \bibinfo{author}{{Jacquemart}, D.}, \bibinfo{author}{{Kleiner}, I.},
  \bibinfo{author}{{Lacome}, N.}, \bibinfo{author}{{Lafferty}, W.J.},
  \bibinfo{author}{{Mandin}, J.Y.}, \bibinfo{author}{{Massie}, S.T.},
  \bibinfo{author}{{Mikhailenko}, S.N.}, \bibinfo{author}{{Miller}, C.E.},
  \bibinfo{author}{{Moazzen-Ahmadi}, N.}, \bibinfo{author}{{Naumenko}, O.V.},
  \bibinfo{author}{{Nikitin}, A.V.}, \bibinfo{author}{{Orphal}, J.},
  \bibinfo{author}{{Perevalov}, V.I.}, \bibinfo{author}{{Perrin}, A.},
  \bibinfo{author}{{Predoi-Cross}, A.}, \bibinfo{author}{{Rinsland}, C.P.},
  \bibinfo{author}{{Rotger}, M.}, \bibinfo{author}{{{\v S}ime{\v c}kov{\'a}},
  M.}, \bibinfo{author}{{Smith}, M.A.H.}, \bibinfo{author}{{Sung}, K.},
  \bibinfo{author}{{Tashkun}, S.A.}, \bibinfo{author}{{Tennyson}, J.},
  \bibinfo{author}{{Toth}, R.A.}, \bibinfo{author}{{Vandaele}, A.C.},
  \bibinfo{author}{{Vander Auwera}, J.}, \bibinfo{year}{2009}.
\newblock \bibinfo{title}{{The HITRAN 2008 molecular spectroscopic database}}.
\newblock \bibinfo{journal}{{J. Quant. Spectr. Rad. Trans.}}
  \bibinfo{volume}{110}, \bibinfo{pages}{533--572}.
\bibitem[{{Teanby} et~al.(2009){Teanby}, {Irwin}, {de Kok}, {Jolly},
  {B{\'e}zard}, {Nixon} and {Calcutt}}]{teanby09a}
\bibinfo{author}{{Teanby}, N.A.}, \bibinfo{author}{{Irwin}, P.G.J.},
  \bibinfo{author}{{de Kok}, R.}, \bibinfo{author}{{Jolly}, A.},
  \bibinfo{author}{{B{\'e}zard}, B.}, \bibinfo{author}{{Nixon}, C.A.},
  \bibinfo{author}{{Calcutt}, S.B.}, \bibinfo{year}{2009}.
\newblock \bibinfo{title}{{Titan's stratospheric C$_{2}$N$_{2}$,
  C$_{3}$H$_{4}$, and C$_{4}$H$_{2}$ abundances from Cassini/CIRS far-infrared
  spectra}}.
\newblock \bibinfo{journal}{{Icarus}} \bibinfo{volume}{202},
  \bibinfo{pages}{620--631}.

\end{thebibliography}
\end{document}